# The electron-ion streaming instabilities driven by drift velocities of the order of electron thermal velocity in a nonmagnetized plasma

Jun Guo [1,2] and Bo Li [3*]

[1] College of Mathematics and Physics, Qingdao University of Science and Technology, Qingdao 266061, China

[2] Key Laboratory of Geospace Environment, University of Science & Technology of China, Chinese Academy of Sciences, Hefei, Anhui 230026, China

[3] Shandong Provincial Key Laboratory of Optical Astronomy and Solar-Terrestrial Environment，Weihai, Shandong 264209, China

---

[*] Author to whom correspondence should be addressed. Electronic mail: bbl@sdu.edu.cn.

We examine the electron-ion streaming instabilities driven by drift velocities of the order of the electron thermal velocity in a nonmagnetized plasma by one-dimensional electrostatic particle-in-cell code which adopts an ion-to-electron mass ratio of 1600. An initial state is set up where the ion bulk speed is zero while the electrons drift relative to ions, and where electrons are much hotter. We examine in detail four runs where drift velocity is systematically varied from lower than to larger than the electron thermal velocity. In all runs the Langmuir waves with Doppler-shifted frequencies dominate early on when streaming instabilities are too weak to discern. And then intense ion-acoustic waves or Buneman instabilities appear, which tend to be accompanied by localized electron and ion beams. Ion-acoustic modes and Buneman modes co-exist in the system when the initial drift velocity is just over the electron thermal speed. Beam modes are excited when the localized beams with large enough velocities appear. In the developed stage of instabilities, the direction in which density depressions propagate is always opposite to that of the localized ion beams. When the initial drift velocity is close to the electron thermal speed, categorizing the relevant instabilities is not easy, and one needs to examine in detail the wave dispersion diagrams at various stages of the evolution of the system.

## 1. Introduction

The electron-ion two-stream instability has been subject to extensive research for a long time [e.g., Bohm et al., 1949; Buneman, 1959; Kellogg et al., 1960; Bernstein et al., 1961]. A collective dissipation mechanism in space plasmas, the behavior of this instability is largely dictated by the relative electron-ion streaming $u_0$ measured in units of the electron thermal speed $v_{the}$. When $u_0/v_{the}$ is sufficiently less than unity, the electron-ion instability belongs to the ion-acoustic type, whereas when $u_0/v_{the}$ far exceeds unity, it turns into the Buneman instability, which becomes unstable when a weak current flows across a plasma [Yoon and Umeda, 2010].

The electron-ion two-stream instability has significant bearings on such important subjects in plasma physics as anomalous resistivity, electron holes and double layers, to name but a few. For instance, in collisionless space plasmas, anomalous resistivity plays an important role for magnetic reconnections. Using a finite difference approximation to the Vlasov-Maxwell equations, Petkaki et al.[2003; 2006] found that the ensemble mean of the ion-acoustic resistivity during the nonlinear regime is higher than estimates at quasi-linear saturation. In magnetic reconnections with a guide field, the Buneman instability produces electron holes, and the associated electron scattering off the holes enhances electron heating in the dissipation region [Che et al., 2009; 2010]. Found in a wide variety of plasmas, double layers are structures wherein ions and electrons are accelerated, decelerated, or reflected by the electric field. They are believed to be a viable mechanism, both observationally [Ergun et al.,2001; 2002; 2009] and theoretically, that can effectively convey the magnetospheric energy in the form of field-aligned currents to auroral particles [Block, 1978]. The necessary condition double layers to form in previous simulations is usually for the electron drift velocity to exceed the electron thermal speed. In other words, double layers are usually attributed to be a result of the Buneman instability. However, observational evidence [Mozer et al., 1977] and theoretical simulations [Sato and Okuda, 1980] exist where double layers along auroral field lines form despite that the electron drift speed is much less than the electron thermal one.

Despite a long history of investigations into the electron-ion streaming instabilities, how the instability transitions from an ion-acoustic type to the Buneman one remains unclear and warrants a dedicated study. Evidently, such a transition will mostly depends on $u_0/v_{the}$, but previous studies suggest that the instability growth rates depend on the electron-to-ion temperature ratio $T_e/T_i$, and the proton-to-electron mass ratio $m_i/m_e$ as well [Treumann and Baumjohann, 1977]. Here the subscripts *e* and *i* denote electrons and ions, respectively. It seems rather strange to note that while most numerical simulation studies adopt a large $T_e/T_i$, there seems to be no study where $u_0/v_{the}$ is systematically varied. Instead, other ingredients tend to be examined. For instance, the seminal paper by Sato and Okuda [1980] adopted a fixed value of $u_0/v_{the} = 0.6$, together with $m_i/m_e = 100$, to offer an in-depth study of the effects of the simulation domain length, thereby showing that double layers can form due to an ion-acoustic instability, as opposed to the traditional Buneman one, if the system is sufficiently long. Adopting values of $u_0/v_{the} = 1.2$ and $m_i/m_e = 25$, Watt et al. [2002] focused their attention on the ion-acoustic resistivity, finding that the ion-acoustic instability may be more important in the magnetopause and low-latitude boundary layer than previously thought. A follow-up study by Hellinger et al. [2004], who adopted $u_0/v_{the} = 1.7$ and $m_i/m_e = 1836$, found that the effective collision frequency of the ion-acoustic instability increases owing to the existence of backward-propagating ion-acoustic waves, which result from induced scattering on protons and contribute to the anomalous electron transport. Furthermore, while most studies work in a frame where initially ions have zero bulk speed and electrons drift relative to ions, Newman et al. [2001; 2002] choose to work in a frame where initially electron and ion distribution functions are both drift Maxwellians with $u_e = v_{the}$ and $u_i = -v_{thi}$, where $u_i$ and $v_{thi}$ are the ion bulk and thermal speeds, respectively. Besides, a mass ratio of $m_i/m_e = 400$ is used. With the introduction of an initial density perturbation, the authors found that a weak density depression in a current-carrying plasma can lead to the formation of a double layer. They further note that double layers, bipolar electric field structures and very-low-frequency whistler emissions may be related.

The present study is intended to be one that systematically varies $u_0$ from less than to larger than $v_{the}$. We note that certainly studies with different $u_0/v_{the}$ exist, but a direct comparison between them is generally not possible since values for mass and/or temperature ratios therein are also different. To facilitate a comparison with available studies, we shall adopt an initial condition where $T_e \gg T_i$. However, a nearly realistic ion-to-electron mass ratio will be used. Furthermore, we will also pay attention to the evolution of electrons and ion distribution functions, which seems not to be a common practice but will offer more clues to the underlying physics. In addition, we will examine how the transition of instabilities is reflected in the electric field and particle behavior.

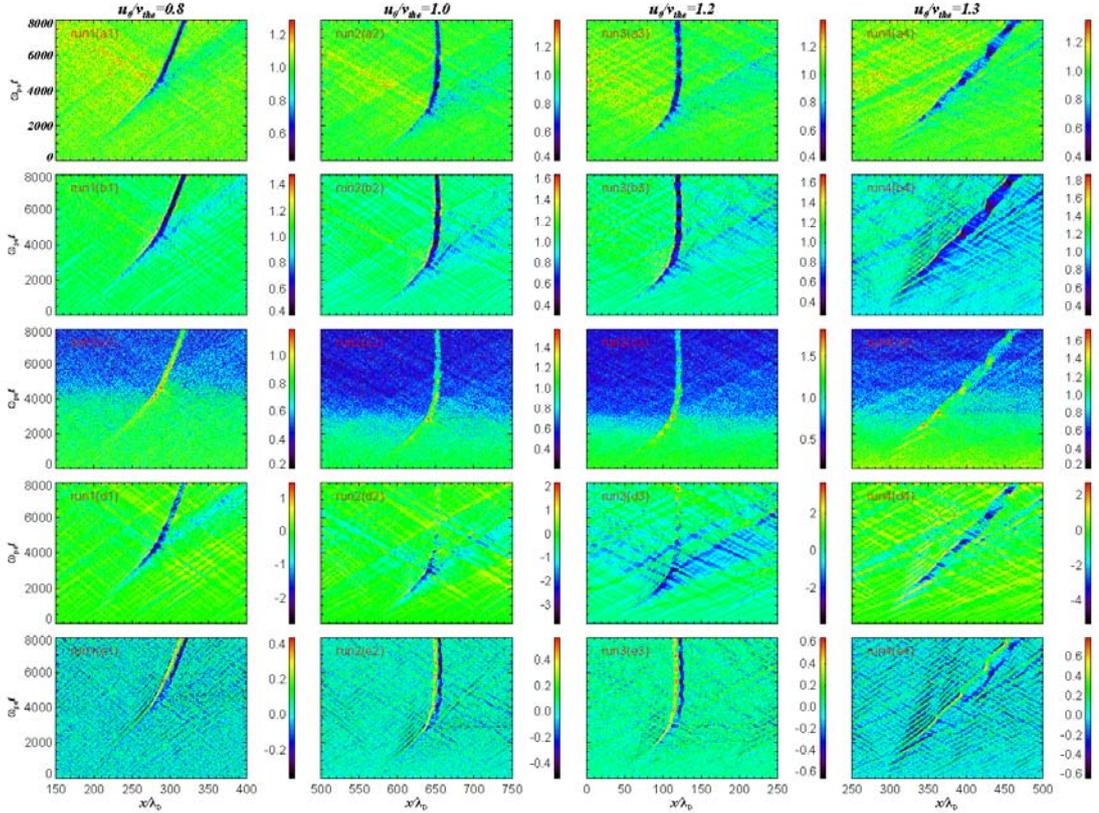

Fig. 1 Time histories of the electron density $n_e/n_0$ (panels a1 to a4), ion density $n_i/n_0$ (b1 to b4), electron drift velocity $u_e/v_{the}$ (c1 to c4), ion drift velocity $u_i/v_{the}$ (d1 to d4), and the electric field (e1 to e4).

## 2. Simulation

A one-dimensional electrostatic particle-in-cell code is used in this study, which neglects the effect of an ambient magnetic field. The electrostatic field is defined on the grids. The periodic boundary condition is used, as was adopted in previous applications of the code [Lu et al. 2004; 2005]. The velocity and time are measured in units of $v_{the}$ and $\omega_{pe}^{-1}$, respectively, with $\omega_{pe}$ being the electron plasma frequency. Likewise, the length in the numerical calculations is in units of the electron Debye length $\lambda_D = v_{the}/\omega_{pe}$. An ion-to-electron mass ratio $m_i/m_e$ of 1600 is chosen. The grid spacing is equal to $\Delta x = \lambda_D$ and the simulation domain length is $L = 1024\lambda_D$. A uniform time step $\omega_{pe}\Delta t = 0.02$ is employed in the simulations. The number of particles in a cell employed for each species of particles is 400. Initially, the ions have no drift velocity whereas electrons drift at $u_0$ relative to ions. Besides, an electron-to-ion temperature ratio $T_e/T_i = 16$ is specified. Four runs are discussed in this paper, differing only in $u_0$. To be specific, runs 1 to 4 correspond to a $u_0/v_{the}$ being 0.8, 1, 1.2, and 1.3, respectively, hence encompassing the range of $u_0$ that is of interest.

### 3. Simulation Results

Figure 1 shows the time histories of the electron density $n_e/n_0$ from (a1) to (a4), ion density $n_i/n_0$ from (b1) to (b4), electron drift velocity $u_0/v_{the}$ from (c1) to (c4), ion drift velocity $u_i/v_{thi}$ from (d1) to (d4), and the electric field from (e1) to (e4). Electron and ion densities with a depression of up to 60% can be seen in the four runs. Besides, while in all runs this depression moves quickly along the *x*- direction in an early stage, its behavior becomes substantially different from one case to another when time further progresses. The density depression in run1(a1) continues to move after $\omega_{pe}t \sim 4600$, albeit with a smaller velocity, due to the decrease of the electron beam velocity. When the initial drift velocities increase, the propagation of the density depression halts in runs 2 and 3 after $\omega_{pe}t \sim 4000$(Figures a2 and a3). As a matter of fact, a close inspection reveals that in these cases after $\omega_{pe}t \sim 7000$ the density depressions start to move in the negative rather than the positive *x*- direction. When the initial drift velocity is further increased to $u_0/v_{the} = 1.3$, as shown in (a4), the density depression almost propagates with the same velocity during the entire simulation. The ion density evolutions are the same as those of electrons except that there is an obvious ion density enhancement located close to the left edge of the density depression.

Figures 1(c1), (c2), (c3) and (c4) suggest that localized electron beams are formed during the simulation, and are located where the density depressions appear. The maximum velocities of electron beams in the four runs are much larger than their initial drift velocities. The velocities of the localized electron beams in the four runs are positive throughout, whereas this is not true for ions. As shown in (d1) and (d4), the ion beams velocities remains negative throughout the simulation. However, while negative early on, the velocities of the ion beams in runs 2 and 3 turn positive from $\omega_{pe}t \sim 7000$ onwards. Actually, running the two runs longer, say, until $\omega_{pe}t = 20000$, we find that both the electron and ion beam velocities remain positive after $\omega_{pe}t \sim 7000$. With the increase of positive ion beam velocities, the density depressions move quickly in the negative *x*-direction from $\omega_{pe}t \sim 9000$ onwards (not shown here). The results of the four runs suggest that the density depression propagation is opposite to the direction of ion beams, suggesting that the direction in which the depression propagates is likely decided by the motion of ions rather than of electrons (for an observed instance of ion beams, please see McFadden et al., 1998).

It is interesting to note that the density depressions as in runs 2 and 3 are nearly stationary during the time interval from $\omega_{pe}t \sim$ 4000 to 7000, as manifested by the ridges in Figs.1(a2) and 1(a3) being nearly vertical. Something similar was also reported by Newman et al. [2001; 2002], where the depression stopped propagating for a shorter period than in our case. This difference is likely due to the different setup between the two simulations, such as the mass ratio, temperature ratio, and boundary conditions. In particular, keeping everything else unchanged, but running run 2 with a smaller mass ratio $m_i/m_e = 400$, we found that the time interval in which the density depression is apparently stationary indeed becomes shorter.

Figures 1 (e1), (e2), (e3) and (e4) indicate that the electric fields have both unipolar and bipolar structures. Evidently, the maximal electric field increases with the increase of the initial drift velocity (note that different scales are used in the colorbars).

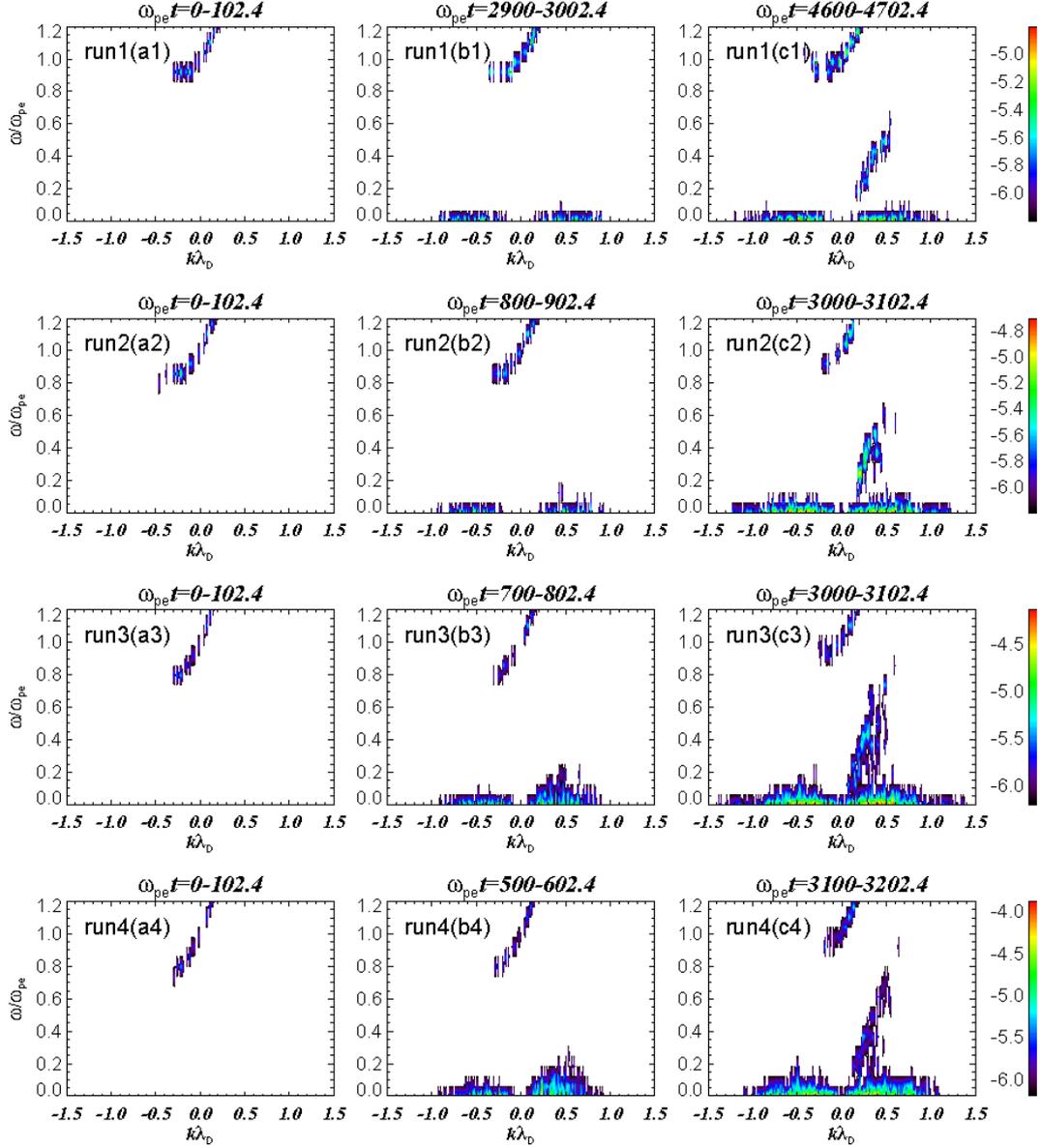

Fig. 2 Dispersion diagrams obtained by Fourier transforming the electric fields $E_x$.

The particle and electric field behavior in the runs as described above actually is largely in line with the typical scenario for the Buneman instability [Buneman,1959]. This is best illustrated if one specializes to the steady state, in which case to prevent charge accumulation, the current density in a 1D system must be the same everywhere. When it passes through a region of decreased electron density, the electrons must be accelerated into this cavity in view of the lowered density of charge carriers. Likewise the ions are accelerated out of the cavity, amplifying the density perturbation. However, this feature is not specific to the Buneman instability. In fact, it is not even present in run 1. This actually lends support to the point we are to make, which is that examining the particle and electric field behavior alone is not sufficient for one to tell what modes of instabilities are operating.

Figure 2 presents the $\omega - k$ diagrams obtained by Fourier transforming the electric field $E_x$ obtained in the four runs. The diagrams are grouped into different time intervals, only three representative ones among which are shown. In run 1, Langmuir waves (LWs, the ridges reaching the top of the panels) with the Doppler-shifted frequency are first excited from $\omega_{pe}t =0$ to 102.4. Compared with the LWs, ion-acoustic waves are too weak to be seen. As time proceeds, ion-acoustic waves appear (panel b1) and intensify (panel c1). In addition, besides the Langmuir waves and ion-acoustic waves, beam-driven modes can be clearly seen in panel c1. Recall that when the beam density is much less than the total density, the dispersion relation of the beam-driven wave is approximately $\omega \approx ku_d$, where $u_d$ is the electron beam drift velocity [Lu et al., 2005]. The slope of the $\omega - k$ diagram shown in Fig.2(c1) is about $1.2v_{the}$, which agrees with the electron beam shown in Fig. 1(c1).

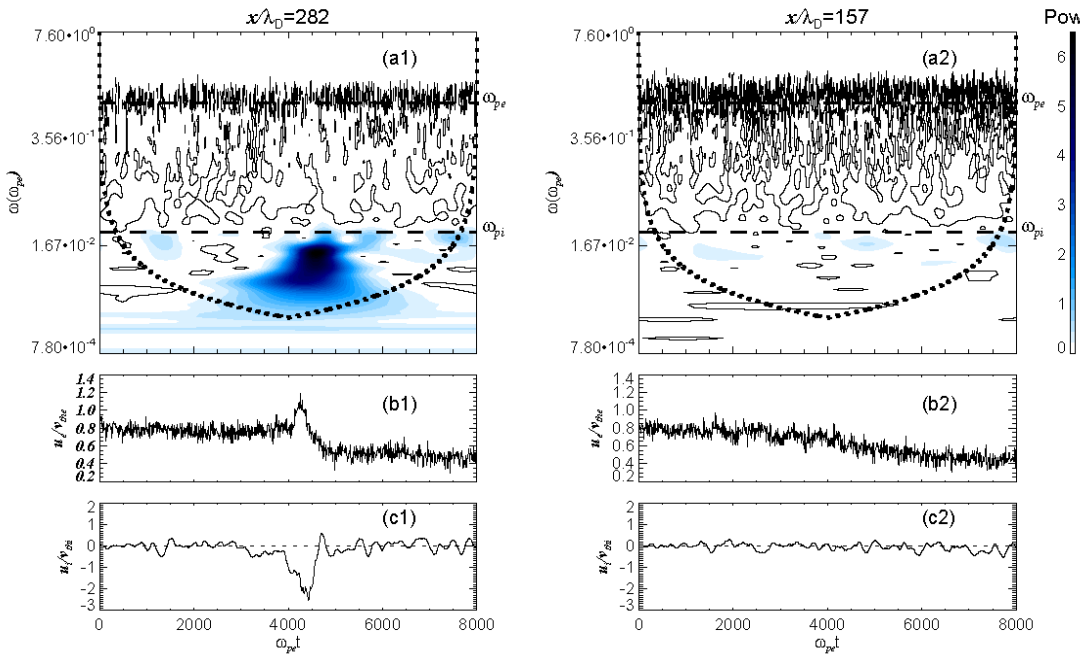

Fig. 3 Evolution of the wavelet power spectra (a1 and a2), electron drift velocity (b1 and b2), and ion drift velocity (c1 and c2) at two different locations $x/\lambda_D =282$ and 157 in run 1.

The evolution of instabilities in run 2 is nearly the same as in run 1, the most intense mode still being the ion-acoustic wave. Compared with run 1, these low frequency instabilities appear earlier due to a larger initial drift velocity. As for run 3 with $u_0/v_{the} = 1.2$, the modes in (b3) propagating in the negative *x*-direction can still be regarded as ion-acoustic waves, while those propagating in the positive *x*-direction should be called the Buneman instability (During this period, electron phase-space holes and plateau distributions form, to be discussed shortly). During the period $\omega_{pe}t$ from 3000 to 3102.4 (c3), the modes propagating in the negative *x*-direction have also turned into the Buneman instabilities. As shown in figures 1(c3) and (d3), the electron beam is accelerated along the positive *x*-direction while the ion beam is accelerated in the opposite direction, resulting an increase of the relative drift velocity between them. This evolution is propitious to the development of Buneman instability. Moreover, the beam mode is clearly present. When it comes to the $\omega - k$ diagrams for run 4 with $u_0/v_{the} = 1.3$, they are close to what one finds in run 3. Of course, the instabilities are now more intense. Figure 2 suggests that the amplitude of the Langmuir waves increases with time. This turns out to be a result of particle trapping, by which a counter-streaming population is generated, exciting in turn electron-electron two-stream instabilities [Jain et al., 2011]. Note that if the initial drift speed is substantially larger than what we adopt here, say $u_0/v_{the} = 4$[Guo & Yu, 2012] or 5 [Jain et al.2011], then the Buneman instabilities are excited first while the LWs are excited afterwards.

Interestingly, the $\omega - k$ diagrams appear different for runs 2 and 3, despite that the characters of particle density, drift velocity and electric field are nearly the same as shown in Fig.1. Likewise, the $\omega - k$ diagrams for runs 3 and 4 are close to one another, whereas Fig.1 suggests that the corresponding time history of the electric field and density depression is very different. From this we conclude that when the initial drift speeds of the electrons, $u_0$, relative to the ions are close to the electron thermal speed $v_{the}$, one cannot safely say that a ratio of $u_0/v_{the}$ that is less than unity will result in the ion-acoustic instability; whereas when the opposite is true the Buneman instability results. To tell what instabilities are present, one at least needs the aid of an omega-k diagram.

Now let us take a closer look at the wave excitation processes. Figure 3 presents the wavelet power spectrum (a1) and (a2), the electron (b1 and b2) and ion drift velocity evolution (c1 and c2) at two different locations in run 1, $x/\lambda_D = 282$ and 157 to be specific. Consider $x/\lambda_D = 282$ first. Evidently, the intense ion-acoustic waves are always accompanied by electron and ion beams. The electron velocity increases first and then the waves are excited. Newman et al. [2001; 2002] pointed out that the resulting small charge separation is that of an incipient double layer, with an electric filed pointing toward the left and thus capable of further accelerating electrons in the direction they were originally drifting. The energy of waves will increase dramatically when the velocity of electron and ion beams abruptly decreases due to wave-particle interaction, which is consistent with observations [Ergun et al., 2001]. Considering $x/\lambda_D = 157$, one can see that although the electron drift velocity begins to decrease at $\omega_{pe}t \sim 4000$(Fig.3(b2)), the instabilities are very weak because no electron and ion beams with high velocities form.

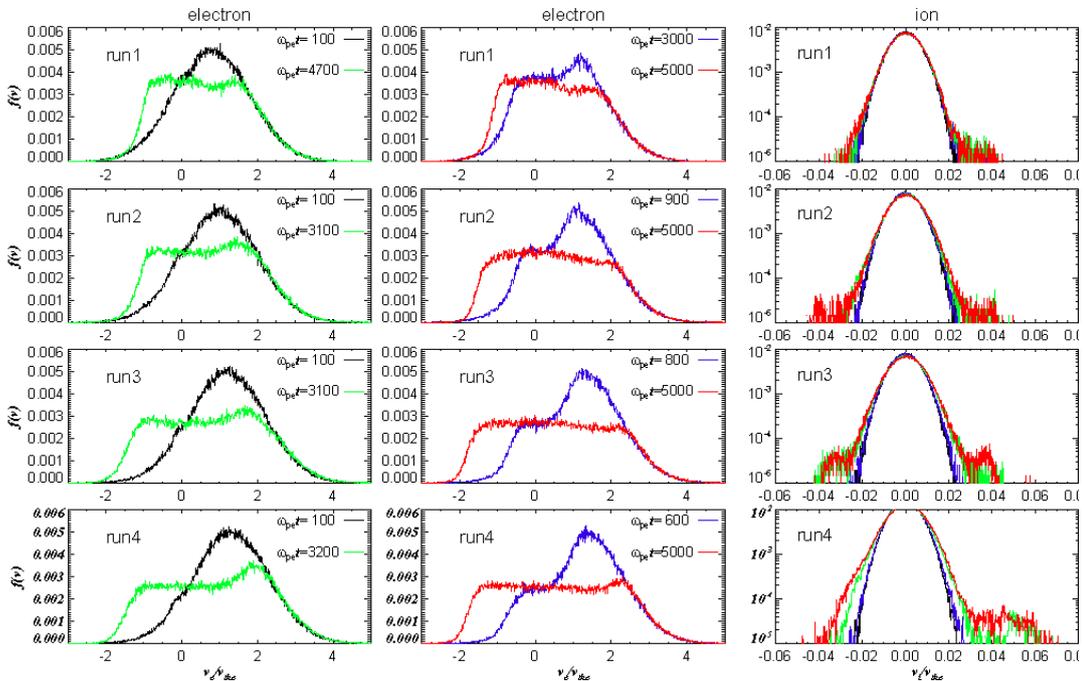

Fig. 4 Electron and ion velocity distribution functions for the four runs.

Figure 4 shows the electron and ion velocity distribution functions for the four runs. Note that a logarithmic scale is used for ions. Consider first the electron distributions. The plateau formation of electrons can clearly be seen at $\omega_{pe}t = 100$ (the black lines in the left panels). However, at this point the ion-acoustic waves or Buneman instabilities are too weak to be seen as shown in Fig.2. Only when the electron plateau has developed sufficiently (the blue lines) do the ion-acoustic waves or Buneman instabilities appear. When a double-peak form is appropriate for describing the electron distributions (the blue and green lines), the left peak is attained at some negative electron speed that is far less in magnitude than the positive speed where the right peak is located. This feature can explain why the modes in the left half of Fig.2 are ion-acoustic modes whereas those in the right half are the Buneman modes. Hellinger et al.[2004] pointed that a pronounced plateau of electron distribution function extending to the negative velocities is a direct consequence of the presence of back-propagating ion-acoustic waves. The red lines present the electron distributions at $\omega_{pe}t = 5000$. At this time, only the distribution for run 4 can be seen as comprising a plateau and a peak.

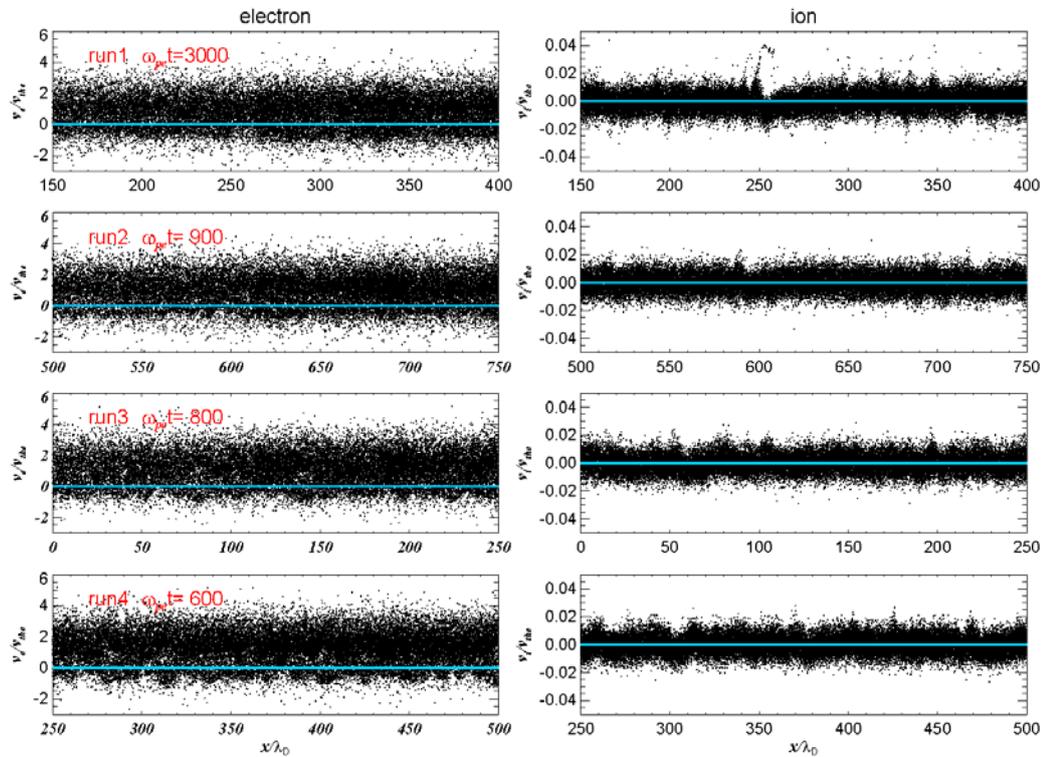

Fig. 5 Electron and ion phase-space structure at different times.

The evolution of ion distributions in the four runs is presented in the right panels. Evidently, the broadening of the ion distribution, i.e., ion heating, is more obvious in run 4 than in run 1. However, ion distributions in runs 2 and 3 at $\omega_{pe}t = 5000$ are very different from those in runs 1 and 4. The red line in run1 suggests that the distribution of accelerated ions is asymmetric about $v_e/v_{the} = 0$. Then it tends to be more symmetric with the increase of the initial drift velocity, as can be seen in run 3. There are two peak values at about $v_e/v_{the} = \pm 0.035$, which means that counter-streaming ion beams are produced. If fact, these ions with high velocities are mainly located in the cavity. These ion beams will counteract the effect of electron beams. Then the relative drift velocity between electrons and ions will decrease substantially. We note that the particle distributions in Fig.4 need not to be confused with those in figure 1, because the distributions herein do not have information regarding the particle positions.

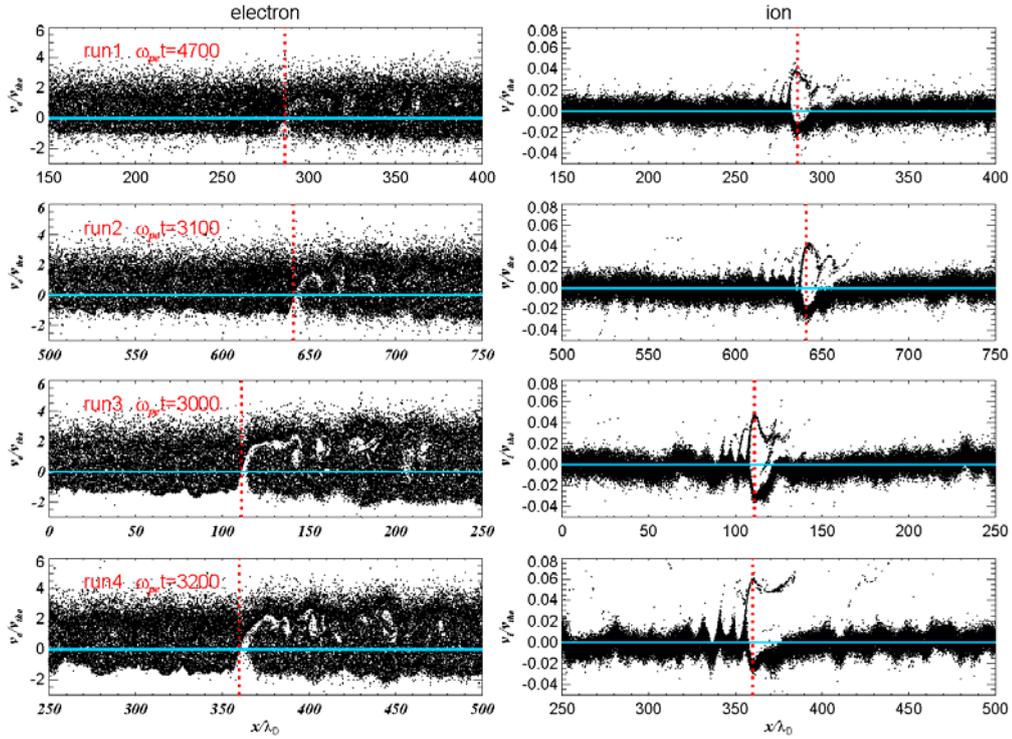

Fig. 6 Electron and ion phase-space structure at different times for the four runs.

Shown in figure 5 are the electron and ion phase-space structures at the time when only the LWs and electron-ion instabilities are found for the four runs. Compared with the results in run 1, the electron phase-space structures in runs 3 and 4 involve the formation of phase-space vortices, consistent with previous studies [Lu et al. 2008; 2010; Wu et al. 2010; 2011]. The larger the initial drift velocity, the more obvious the phase-space vortices. We note that these vortices will be clearer still if a much larger value is adopted for the initial relative drift [Jain et al., 2011; Yoon et al., 2010]. Concerning the ions, one may see that obvious ion holes appear only in run 1 with $u_0/v_{the} < 1.0$.

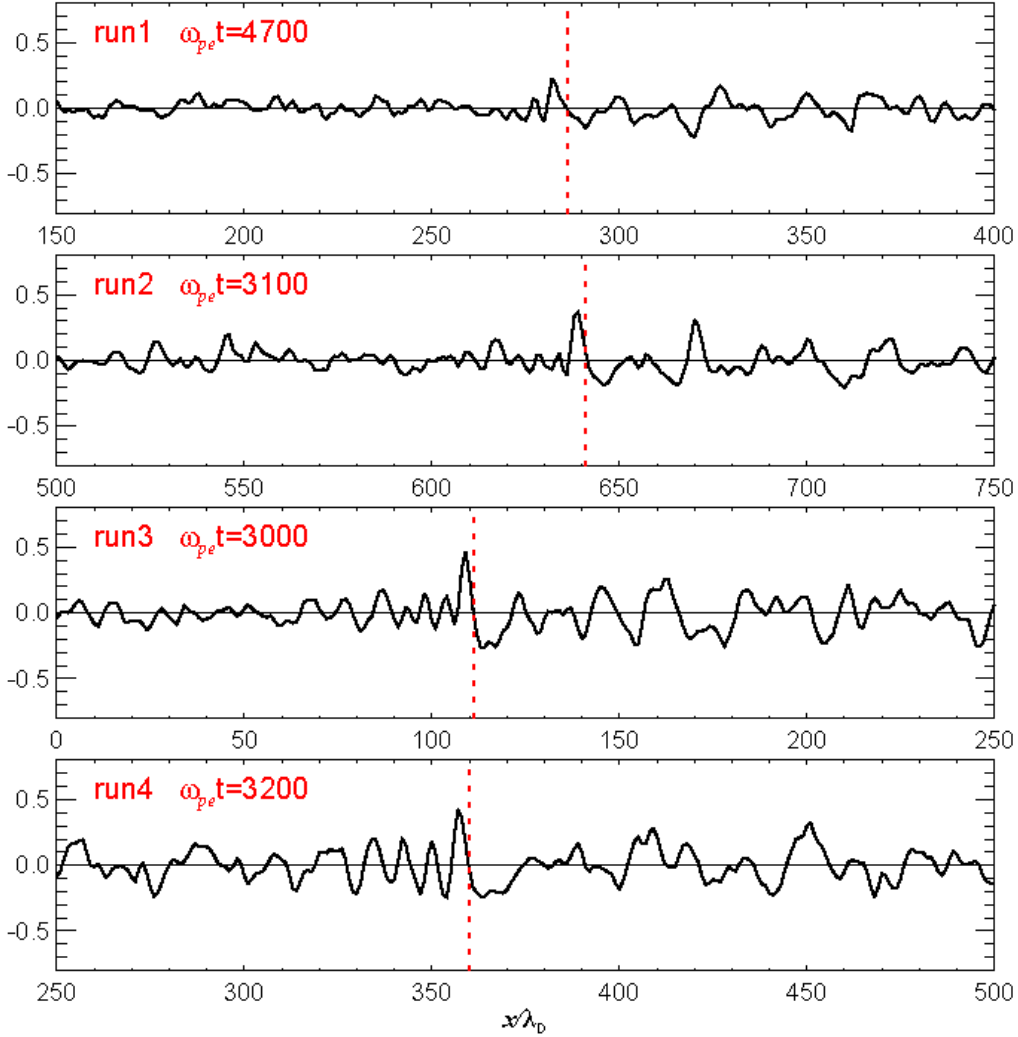

Fig. 7 Spatial distribution of electric fields corresponding to the distributions in Fig.6.

Shown in Fig.6 are the electron and ion phase-space structures at the time when the beam modes are also found for the four runs. While in Fig.5 the electron phase-space vortices are symmetric about $v_e/v_{the} = 0$, in Fig.6 some holes appear in the electron phase-space. The holes start with negative electron speeds and extend into the positive portion of the electron speeds (see the vertical plume-like features indicated by the vertical red dotted lines). In addition to this component that is almost fixed at some distance but occupies a finite volume in the velocity space, in runs 3 and 4 the holes also comprise a component that is horizontally directed, corresponding to an almost fixed speed but a finite volume in the configuration space. In a sense, this horizontal feature is opposite to what we expect for an electron beam. These structures agree with previous works [Newman et al., 2001; 2002; Goldman et al., 2003]. In light of the diagram presented in Fig.3, these holes should be associated with the beam modes and electron-ion two-stream instabilities. We note that in the work by Newman et al.[2001], the authors introduced a charge-neutral density depression into their background as an initial perturbation. One may therefore conclude that the initial "seed" depression is not a prerequisite for electron phase-space holes to occur.

Shown in Fig.7 are the spatial distributions of the electric field recorded at the same time as the distributions given in Fig.6 for the four runs. Unipolar and bipolar electric field structures can be seen to co-exist. The unipolar negative electric field of the double layer is clearly visible where density depressions and beams are located, as shown in Fig.6. The double layer spans about 10 Debye lengths ($\lambda_D$), in agreement with observations [Ergun et al., 2001]. The electric field close to the left edge of the double layer is quite turbulent, as was pointed out by [Newman et al., 2001; 2002]. Compared with the distributions in Fig.6, these turbulent electric fields show a clear one-to-one correspondence with the features seen in the ion distributions, where the ion holes are semicircular in phase-space when the strong double layers appear.

The Fast Auroral SnapshoT (FAST) measurements indicate that double layers may remain intact for at least $\sim 10\text{ms} \approx 4 \times 10^3 \omega_{pe}^{-1}$ [Ergun et al., 2001]. This lifetime is significantly longer than what is found for the double layers in our simulation ($\approx 1 \times 10^3 \omega_{pe}^{-1}$) or $\sim 700 \omega_{pe}^{-1}$ as reported by Newman et al.[2001; 2002]. Note that Newman et al.[2001] attributed this discrepancy to the low ion-to-electron mass ratio adopted therein ($m_i/m_e = 400$). Adopting a value for $m_i/m_e$ as large as

1600 does bring the numerically computed lifespan of the double layers closer to the observed ones, but does not seem to have resolved this issue. In addition, the unipolar and bipolar electric fields appear alternately in run 2 where $u_0/v_{the} = 1.0$. What gives rise to this is certainly worth study but is beyond the scope of the present manuscript.

**4. Conclusion and Discussion**

This study is motivated by the apparent lack of a study that is dedicated to examining the behavior of electron-ion streaming instabilities when the initial drift speeds of electrons relative to ions are of the order of the electron thermal speed $v_{the}$. To be specific, using a 1-D particle-in-cell code with periodical boundary conditions, we systematically varied $u_0$ from lower than to larger than $v_{the}$, and examined in detail how the system comprising particles, electric fields, and waves evolves from an initial state where the electron temperature far exceeds the ion one ($T_e/T_i = 16$). Besides, we adopted an ion-to-electron mass ratio $m_i/m_e$ of 1600.

The results from our computations can be summarized as follows.

1) In the presence of an initial relative drift $u_0$ between electrons and ions, streaming instabilities develop but follow different paths with different $u_0$. In the early stage, the Langmuir waves dominate. Then some intense ion-acoustic waves or Buneman instabilities appear, with the instabilities being significantly amplified when strong electron and ion beams form.

2) Categorizing the instability is not an easy matter, and certainly cannot be done a priori by following the rule of thumb that when $u_0$ is less than $v_{the}$, the ion-acoustic instability results, whereas when the opposite is true one expects to see the Buneman instability. At least one needs to examine in detail the $\omega - k$ diagrams at various stages of the evolution of the system.

3) When the initial relative drift $u_0$ almost equals the electron thermal speed ($v_{the}$), something interesting happens in that the density depression will develop, then stay motionless for $\sim 3000\omega_{pe}^{-1}$, after which its propagation direction reverses. This reversal happens in conjunction with the ion beam developing a substantial velocity in the direction opposite to the electron beam. This suggests that the propagation direction of density depression in the developed stage of the instabilities may be determined by the ions rather than the electrons.

**Acknowledgments**

This research was supported by the National Natural Science Foundation of China under Grant No. 41204115, 41174154, the Provincial Natural Science Foundation of Shandong via grant JQ201212, and by the Key Laboratory of Geospace Environment, University of Science & Technology of China, Chinese Academy of Sciences